\documentstyle[epsfig]{europhys}

\newif\ifboo \boofalse

\def\Review#1{\boofalse{\it #1},}

\def\Vol#1{\ifboo Vol. {\bf #1}\else{\bf #1}\fi}
\def\Year#1{\ifboo #1\else(#1)\fi}

\def\Page#1{\ifboo {\rm p. #1}\else{\rm #1}\fi}

\begin{document}

\euro{}{}{}{}

\Date{}


\title{Anti-Localisation to Strong Localisation:
The Interplay of Magnetic Scattering and Structural Disorder}

\author{Sanjeev Kumar and Pinaki Majumdar}

\institute{ Harish-Chandra  Research Institute,\\
 Chhatnag Road, Jhusi, Allahabad 211 019, India }

\rec{}{}


\maketitle

\begin{abstract}

We study the effect of magnetic scattering on transport in a system with strong
structural disorder, using exact finite size calculation of the low frequency
optical conductivity. At weak electron-spin coupling, spin disorder leads to a
{\it decrease}  in resistivity, by weakening the quantum interference precursors
to Anderson localisation. However, at strong electron-spin coupling, the `double
exchange' limit, magnetic scattering increases the effective disorder, sharply 
{\it increasing} the resistivity.  We illustrate the several unusual transport 
regimes in this strong disorder problem, identify a re-entrant 
insulator-metal-insulator transition, and map out the phase diagram at a 
generic electron 
density.

\end{abstract}

\section{Introduction}
The physics of transport and localisation in the presence of structural
disorder has been extensively studied  \cite{lee-ram}.
In three dimension (3d) increasing disorder leads to a monotonic 
increase in the fraction of localised states, and the resistivity in the 
extended regime, leading finally to the Anderson metal-insulator 
transition (MIT).

The interplay of scattering from `paramagnetic' moments 
with scattering from structural disorder leads to several new 
transport regimes, whose character is poorly understood. 
The presence of weak magnetic scattering actually {\it weakens} the
localising effect of disorder, as discovered by Lee \cite{lee-wl-sf}
and by Hikami {\it et al.} \cite{hik-wl-sf}, while 
strong magnetic coupling, the `double exchange' limit, {\it enhances}
the localising effect of structural disorder \cite{li-dex-dis}. 
There is no understanding of how these two endpoints are connected.
This experimentally relevant  ``middle'' is wide, 
unexplored, and beyond the reach of
standard Boltzmann transport theory \cite{ziman}.
In this paper we present essentially exact results on the 
conductivity and MIT  
considering the combined effect of structural
disorder and magnetic scattering, and  
illustrate the novel 
transport regimes in the problem. 

An understanding of transport properties of disordered magnetic 
systems is
of particular relevance now because of intense experimental activity in
diluted magnetic semiconductors \cite{dms1,dms2,dms3}
 (DMS), amorphous magnetic semiconductors \cite{gdsi1,gdsi2},
{\it e.g} GdSi, and the manganites \cite{cmr-ref}. 
Some of these systems, notably those where the magnetic coupling arises
from Hunds rule, as in $d$ electron systems, require an understanding
beyond the `weak magnetic scattering' studies in the localisation 
literature. A quick look at the resistivity in 
the paramagnetic phase in these materials reveal that they are
all `poor metals'.
The resistivity at 300K in the DMS, Ga$_{1-x}$Mn$_x$As,
at $x \sim 0.08$, is $\approx 4-6$ m$\Omega$cm \cite{dms3}.
For $a$-GdSi, in its `metallic regime' this is $\sim 3$
m$\Omega$cm \cite{gdsi1}.
For the manganite La$_{1-x}$Sr$_x$MnO$_3$ (LSMO),
at large doping, $x >  0.3$,
where electron-phonon coupling effects are expected to be weak, the
room temperature resistivity is $>  5$ m$\Omega$cm \cite{lsmo}. 
To put these numbers in perspective, let us use the Mott `minimum
metallic conductivity' \cite{mott-mit} as reference. The Mott
`minimum' conductivity  is $\sim 0.03 (e^2/{\hbar a_0})$ and if 
we use a lattice constant $a_0 = 3 \AA$ then $\sigma_{Mott}
\sim 2.5 \times 10^4~(\Omega$m)$^{-1}$, {\it i.e},
the `Mott resistivity' $\rho_{Mott}$ is
approximately $4$ m$\Omega$cm. Comparing with the experimental
data quoted above, for all the systems concerned, 
$\rho/\rho_{Mott} \sim  {\cal O}(1)$  
in the paramagnetic
phase.  There is no standard theory for
analysing the scattering from structural and magnetic disorder in 
this regime.

A complete theory of transport, for example the
temperature dependence,  in any of these
systems would of course have to deal 
with {\it annealed spin disorder}. We will discuss these
effects in the future, but focus here on the simpler case of
scattering from a 
combination of quenched structural and magnetic disorder.
Our principal results are contained in Fig.1, showing the `global'
behaviour of the  
resistivity,  and the phase diagram in
Fig.3.

\section{Model}
We study transport in the following
model:
\begin{equation}
H = -t \sum_{\langle ij \rangle, \sigma }   
c^{\dagger}_{i \sigma}c_{j \sigma}  
+ \sum_{i\sigma} (\epsilon_i - \mu) n_{i \sigma} 
- J'\sum_i {\bf \sigma}_i.{\bf S}_i 
\end{equation}

This is the `Anderson disorder' problem, with the electrons being coupled
additionally to spins, ${\bf S}_i$, through $J'$. We use 
$J' > 0$. 
The hopping is only between nearest neighbour sites in a 
simple cubic lattice.  The random on site potential, 
$\epsilon_i$, is uniformly distributed between $\pm \Delta/2$.
The dimensionless parameters in the problem are 
disorder $\Delta/t$, 
magnetic coupling $J'S/t$ and the electron density $n$,
controlled by $\mu$.
We absorb $S$ in our magnetic coupling $J'$, assuming
$\vert {\bf S}_i \vert = 1$. $W = 12t$ is the bandwidth.
We assume that the spin distribution is uncorrelated between sites,
and isotropic on site, {\it i.e} each spin can point anywhere on the
surface of a unit sphere.
The energies are measured in units of $t$, set to $1$.

\section{  Conductivity calculation}
We estimate the d.c conductivity, $\sigma_{dc}$,
 by using the Kubo-Greenwood 
expression \cite{mahan} for the optical conductivity. 
In a disordered non interacting system: 
\begin{equation}
\sigma ( \omega)
=  {A \over N}
\sum_{\alpha, \beta} (n_{\alpha} - n_{\beta})
{ {\vert f_{\alpha \beta} \vert^2} \over {\epsilon_{\beta} 
- \epsilon_{\alpha}}}
\delta(\omega - (\epsilon_{\beta} - \epsilon_{\alpha}))
\end{equation}
The constant $A= 
 (\pi e^2)/ {\hbar a_0 } $. 
The matrix element $f_{\alpha \beta} = \langle \psi_{\alpha}
\vert j_x \vert \psi_{\beta} \rangle$ and  the current operator
in the tight binding model is
$j_x = i t a_0 e \sum_{i, \sigma} (c^{\dagger}_{{i + x a_0},\sigma}
c_{i, \sigma} - h.c)$. The $\psi_{\alpha}$ etc
 are single particle eigenstates,
for a given realisation of disorder, and $\epsilon_{\alpha},
\epsilon_{\beta}$ etc
are the
corresponding eigenvalues. The   
$n_{\alpha}= \theta(\mu
- \epsilon_{\alpha})$ etc are occupation factors.

The conductivity above is prior to disorder averaging.
We work in a $L_T \times L_T \times L$ geometry and, given the
finite size, the $\delta$ function constraint in $\sigma(\omega)$
cannot be satisfied
for arbitrary $\omega$. 
\begin{figure}

\hspace{3cm}
{\epsfxsize=8.0cm, \epsfysize= 6.8cm, \epsfbox{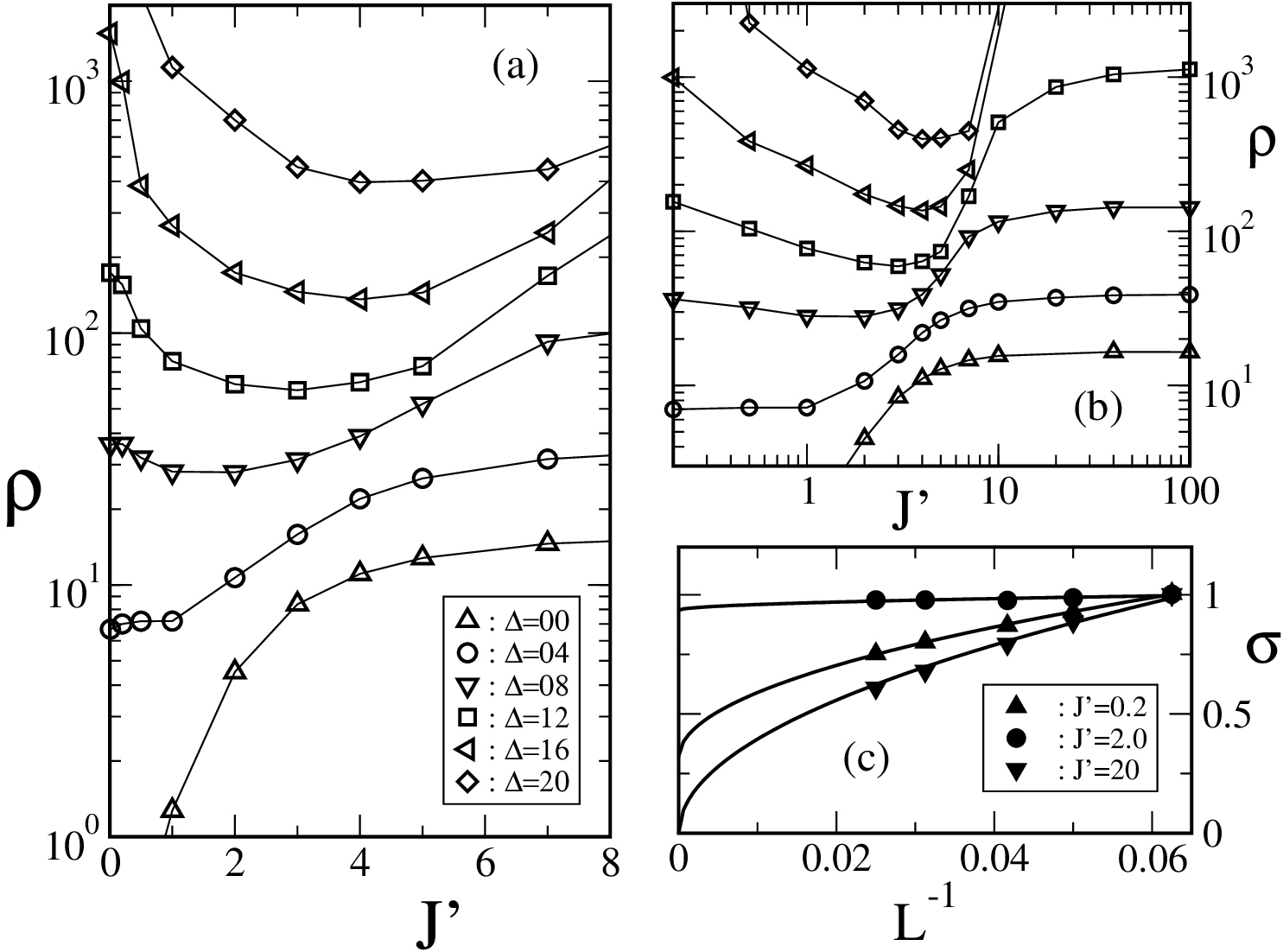}}

\vspace{-.2cm}

\caption{Variation of resistivity, $\rho$, with $J'$ and $\Delta$.
Density $n=0.5$. Data obtained by extrapolation on a sequence of sizes
$6 \times 6 \times L$, with $L$ varying from $24$ to $40$.
Panel $(a)$ shows $\rho$ in the intermediate $J'$ regime, while
panel $(b)$ shows a `global' view, with a logarithmic $J'$ scale.
Panel $(c)$ shows how the `d.c conductivity' is obtained by
extrapolation, illustrating the behaviour for various  $J'$ at 
$\Delta=16$, {\it i.e}, near the Anderson MIT.
The conductivity in panel $(c)$ is normalised to $\sigma(L=24)$.
}

\vspace{-3mm}
\end{figure}
However, we can still calculate the average `low frequency' 
conductivity, defined below, and extract the d.c. conductivity 
by studying a sequence of system sizes.

We set $L_T =6$, and sum over the $\delta$ functions to 
compute the low frequency average
 $ \sigma_{av}(\mu, \Delta \omega, L)
=  (\Delta \omega)^{-1} \int_0^{\Delta \omega}
\sigma(\mu, \omega, L)d \omega
$, for 
$\{L: 24, 28, 32, 36, 40\}$. using  periodic boundary condition in all
directions. The averaging interval is reduced with
increasing $L$, with 
$\Delta \omega \sim B/L$.
The constant $B$ is fixed by setting $\Delta \omega = 0.06$ at
$L=40$.
$\sigma_{av}$ is  disorder averaged  over $N_r$ realisations,
which 
involve `cross averaging' over $\epsilon_i$
and spin configurations, {\it i.e} both the $\epsilon_i$ and ${\bf S}_i$ 
are chosen freshly for each realisation. 
We use $N_r \sim 100$ for the largest
size $(L=40)$, increased to $N_r = 400$ for $L=24$.
Denoting the disorder averaged 
low frequency conductivity for a size $L$
and $\Delta \omega = B/L$, as ${\bar \sigma}_{av}(\mu, B/L, L)$, the 
d.c conductivity is obtained as  
$\sigma_{dc}(\mu) = {lim}_{L \rightarrow \infty} 
{\bar \sigma}_{av}(\mu,
B/L, L)$.
The chemical potential is set to target the required 
electron density $n$.
Our transport calculation method 
and some benchmarks will be  
discussed in detail elsewhere
\cite{sk-pm-long-transp}.
To convert to `real' units, note that our conductivity results are in units
of $(\pi e^2)/{\hbar a_0}$, {\it i.e} $\sigma =1$ on our scale corresponds to
$\approx 10^2 \sigma_{Mott}$.
All our results in this paper are for $n=0.5$, which is `quarter
filling' in the weak $J'$ limit, and half-filling (of the lower
band) at  large $J'$. $n=0.5$ is a generic `high density' 
case and exhibits all the interesting regimes in transport.
The {\it density dependence } of
transport properties will be 
discussed separately \cite{sk-pm-long-transp}.

\section{Transport Regimes and Insulator-Metal Transitions}
The qualitative features in transport are immediately visible 
in the resistivity,  Fig.1, panels $(a)-(b)$, 
the most noteworthy being the
strongly non monotonic behaviour of $\rho(J')$ at large
$\Delta$.
To set the stage for discussing the {\it interplay}
of structural disorder and magnetic scattering, let us
quickly review the {\it individual effects} 
arising from each of them.

Analytic approaches to the structural disorder problem,
beyond Born scattering,
consider the weak localisation (Cooperon) contributions to the 
conductivity in a $(k_Fl)^{-1}$ expansion, 
\begin{figure}

\hspace{.5cm}
{\epsfxsize=6.0cm, \epsfysize= 7cm, \epsfbox{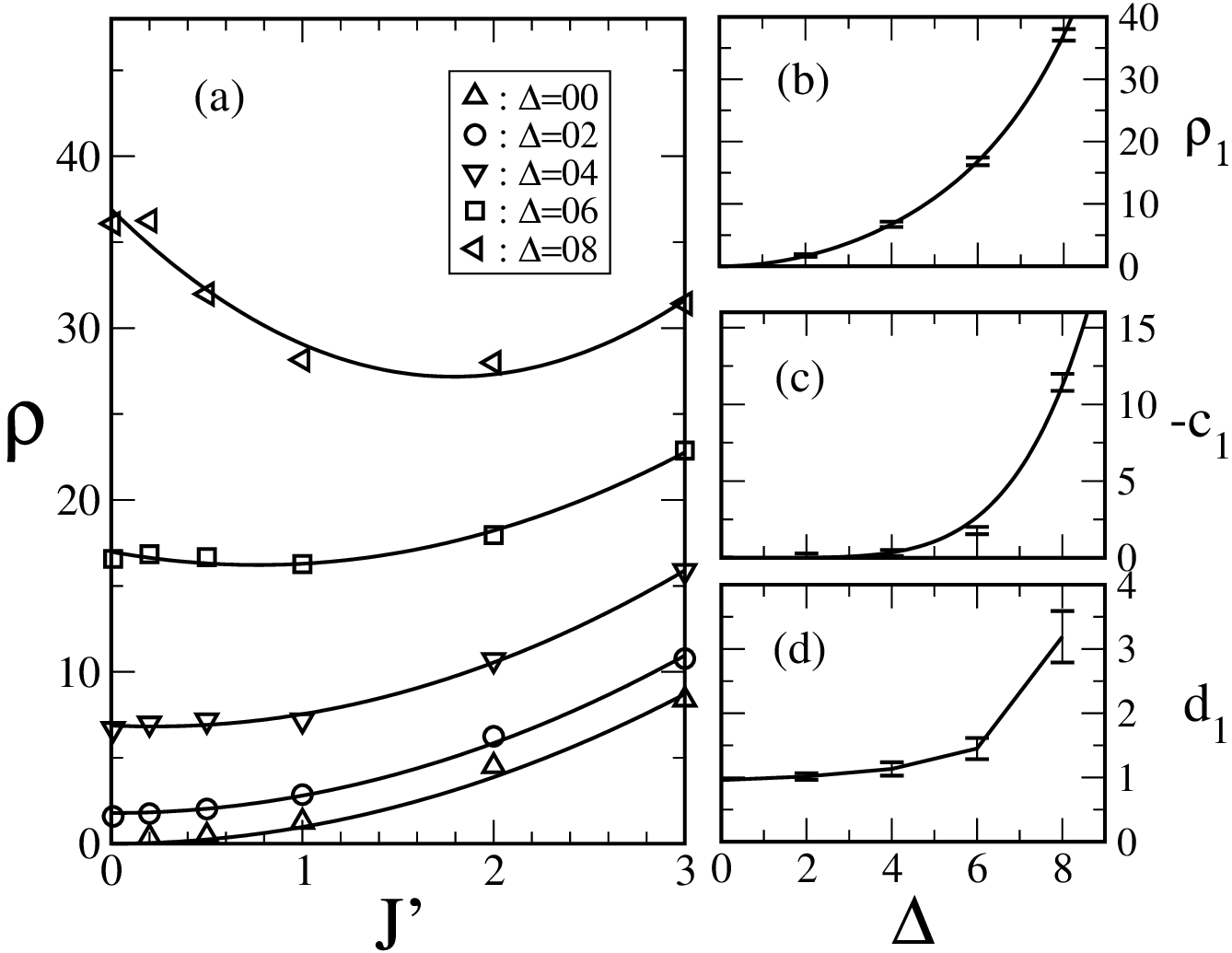}}
\hspace{-.1cm}
{\epsfxsize=6.0cm, \epsfysize= 7cm, \epsfbox{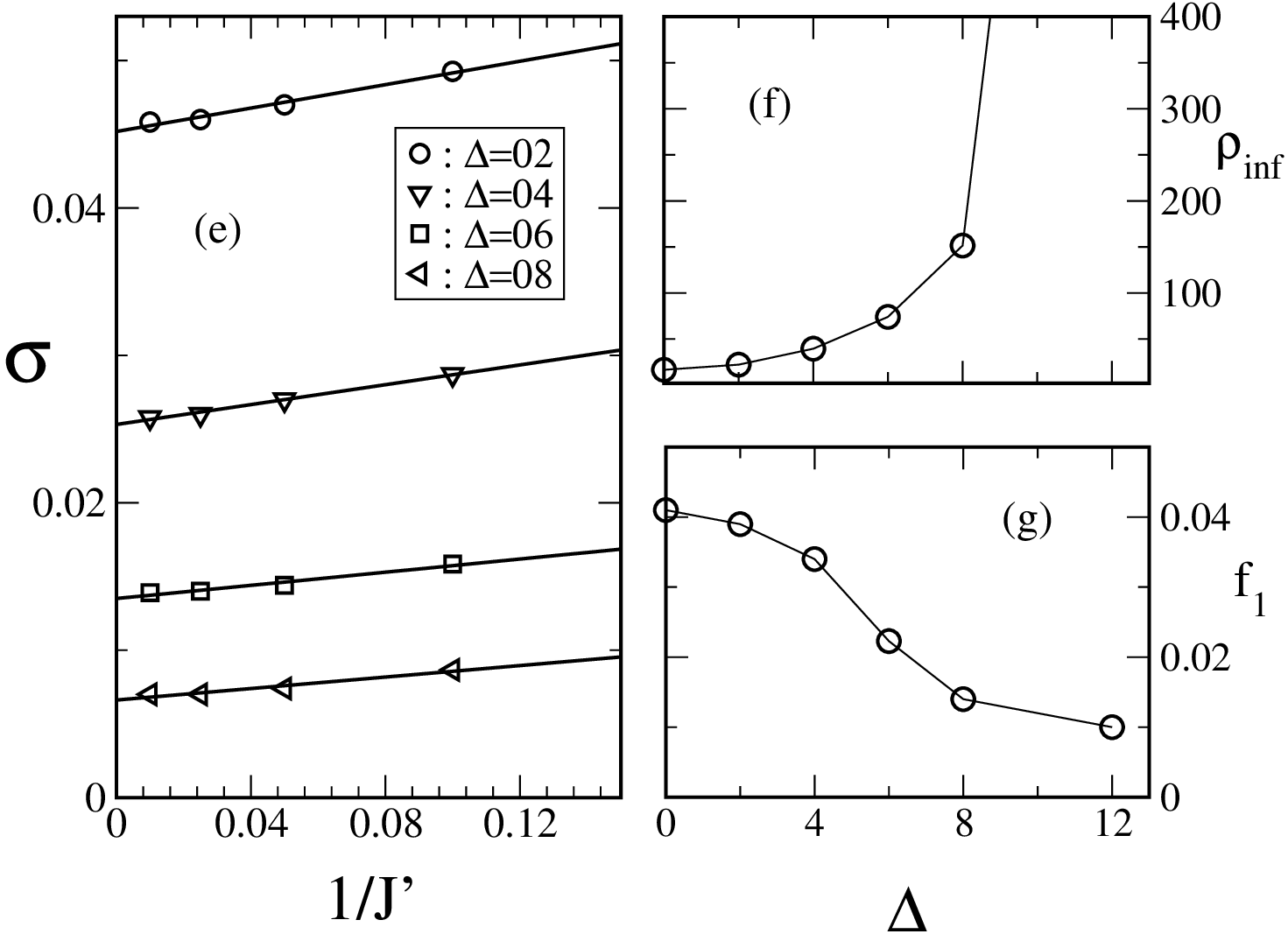}}

\caption{Left Fig: Resistivity at small $J'$:
$(a)$~Quadratic fit to the extrapolated resistivity: $\rho(J')
\approx  \rho_1( \Delta) +    c_1( \Delta) \vert J' \vert + 
d_1 (\Delta) J'^2 $,
 at weak coupling, and moderate $\Delta$.
$(b)$~$\rho_1(\Delta)$, with a polynomial fit (see text),
$(c)$~The coefficient $-c_1$, with a power law fit (see text),
and
$(d)$~The coefficient $d_1$.
Right Fig:  Resistivity at  large $J'$:  
$(e)$~Fit to $\sigma(J', \Delta)$ of the form $\sigma(\infty, \Delta)
+ f_1/J'$. $(f)$~Resistivity in the double exchange limit
$\rho(\infty, \Delta) = 1/\sigma(\infty, \Delta)$. $(g)$~The
coefficient $f_1$.}
\vspace{-2mm}

\end{figure}
with $k_F$ and $l$ being the
Fermi wavevector and mean free path respectively.
Beyond this one can
use the approximate self-consistent theory (SCT) of localisation
\cite{sct},
or depend on numerical results. 
With growing disorder the resistivity, at fixed electron density,
increases monotonically, finally diverging at the Anderson transition.

For magnetic disorder,
the lowest order
effect arises from  Born
scattering, with $\rho(J') \propto J'^2$. Beyond this regime
the resistivity shows a dependence \cite{sk-pm-long-transp}
of the form 
$\rho(n, J') \sim b_1(n) J'^2 + b_2(n) J'^4$, 
upto moderate coupling, $J'/t \sim 3$, with  
$b_1, b_2$ being 
electron density dependent coefficients, 
At strong coupling, and the 
band center, $\rho(J')$ saturates with growing $J'$.

Our interest is in the  combined effect 
of these two scattering mechanisms.
There are tentatively {\it five} different 
transport regimes in the
problem, and, except for $(i)$ below, 
all of them are beyond the 
reach of standard transport theory. These are:
$(i)$~the weak scattering regime, where  Mathiessens rule
holds, $(ii)$~spin flip correction to weak localisation (WL),
with the $\Delta$ dependence showing WL
corrections and
spin flip increasing the conductivity, 
$(iii)$~spin dephasing driven insulator-metal transition (IMT), 
occuring over a window in
$\Delta$, $(iv)$~the disordered double exchange (DE) limit,
and $(v)$~the intermediate coupling `metal'.
We analyse our data in the spirit of this classification.

$(i)$~{\it Weak scattering, obeying Mathiessens rule:}
The regime of weak structural disorder and weak electron-spin coupling can
be understood in terms of
additive Born scattering,  with the net 
scattering rate, $\Gamma(\Delta, J') \approx a_1\Delta^2 + b_1 J'^2 $.
The resistivity is additive and both contributions are 
described by
lowest order perturbation theory.
This corresponds to the bottom left hand corner in Fig.1.(a).
The `parallel' curves in Fig.2(a) for $\Delta$ upto $4$, and 
the regime $J' 
\le 3$ broadly identify the domain of 
`Mathiessens rule'.
The net resistivity in this regime, Fig.2.$(a)$, is $ <  0.1 
\rho_{Mott}$,
{\it i.e}, a few hundred $\mu \Omega$cm.

$(ii)$~{\it Spin flip correction to weak localisation:}
In the structural disorder problem 
the quantum
corrections to the conductivity become important
with growing $\Delta$, 
and show up 
via the WL effect. This effect is already beyond
standard transport theory. The leading 
corrections to $\sigma(\Delta)$ in three dimension, 
beyond Boltzmann transport, have  been worked out \cite{bel-kirk-3d}.
Exact numerical calculations \cite{sk-pm-long-transp,nik-allen}
 suggest that the Boltzmann result and the low order quantum corrections
describe the resistivity upto $\Delta/W \approx 1$.
In this regime, with $\Delta/W \sim 1$, the effect
of magnetic scattering,  with $J'/W \ll 1$ is non trivial.  
Spin flip scattering of the electrons 
\cite{lee-wl-sf,hik-wl-sf}, 
by the random magnetic moments serves to {\it suppress} the 
localising effect of structural disorder,  Fig.1.(a)  and Fig.2.(a),
as we discuss below.

Just as inelastic scattering leads to decoherence, and a 
suppression of quantum interference, spin
flip scattering of the electrons, arising from
processes of the form $(S_{xi} + i S_{yi})
c^{\dagger}_{i \downarrow} c_{i \uparrow}$ 
etc, leads to {\it spin decoherence}
and a corresponding cutoff to WL  corrections. 
If the dephasing time is $\tau_s$, then the spin diffusion  length is
$l_s = \sqrt {D  \tau_s}$, with the diffusion constant 
$D \propto \Delta^{-2}$, and
$\tau_s^{-1} \propto J'^2$. The WL correction gets {\it
corrected} \cite{lee-ram} 
by $\delta \sigma \propto  l_s^{-1} \propto \Delta \vert J \vert$. 
If the WL  correction to the conductivity scales as
$\sim (k_Fl)^{-1} \sim \Delta^2$, then the corresponding spin flip
correction to the {\it resistivity}
should  behave as $\delta \rho \sim -\Delta^5
\vert J' \vert$.

Fig.2.(a)  shows $\rho(J', \Delta)$ in the
moderate $J'$ region, with $\Delta$ increasing beyond the 
Born scattering regime.
At  weak disorder the resistivity has the expected  form 
$\rho(\Delta, J') \propto a_1 \Delta^2 
+ b_1 J'^2$, while at larger disorder we expect the $\Delta$
dependence to be stronger, {\it and simultaneously a positive 
contribution to the conductivity $\delta \sigma  
\propto  \vert J' \vert$ to show up}. The coefficient of the
$J'^2$ term could also be renormalised at large $\Delta$.
Taking these possibilities into account we  fit 
the low $J'$ resistivity to the form:
$
\rho( \Delta, J')\vert_{J' \rightarrow 0}
 \approx  
\rho_1( \Delta) +  b_1 J'^2 + c_1( \Delta) 
\vert J' \vert + 
f( \Delta)J'^2 $,
where the coefficients would depend on electron density.
$\rho_1(\Delta)$ tracks the localising effect of structural disorder,
$c_1(\Delta)$ measures the `antilocalising' effect of spin flip scattering,
and $d_1 = b_1 + f(\Delta)$  monitors the renormalised `Born scattering' from spins.

For $n=0.5$, the resistivity from
structural disorder, $\rho_1(\Delta) \approx a_1 \Delta^2 + a_2 
\Delta^4 + a_3 \Delta^6$, as we fit in Fig.2.(b), with 
$a_1 = 0.408$, $a_2 = 3 \times 10^{-4}$, and $a_3
= 4 \times 10^{-5}$. The coefficient of the `anti-localisation'
term, Fig.2.(c) is $c_1(\Delta)
 \approx c_0 \Delta^{\alpha}$, with $c_0 = -3.4
\times 10^{-4}$, and $\alpha \sim 5$. The (renormalised)
coefficient of the $J'^2$ term, $d_1 = b_1 + f(\Delta)$,
has the form indicated in
Fig.2.(d). 
While Fig.2.(b) highlights the rapid growth in 
resistivity due to structural disorder, Fig.2.(c) indicates
how this growth is cutoff via the spin flip scattering effected
by $J'$. This affects the metal-insulator phase boundary 
as we discuss next.

$(iii)$~{\it Spin dephasing driven IMT:}
There is no perturbative scheme as the disorder increases beyond the
weak localisation regime. The SCT provides a guide and 
of course there is extensive numerical work on the Anderson transition.
Unfortunately, the SCT has no equivalent that includes magnetic
scattering as well (although there were early attempts in two
dimension \cite{yoshioka}), and we do not know of even 
numerical work probing the MIT including magnetic scattering.

As $\Delta$ drives the system towards localisation,
the effect of {\it weak} $J'$ is dramatic. 
The small increase
in conductivity, correcting the weak localisation 
contribution, now
develops into a full fledged anti-localisation effect, 
enlarging the domain of the metallic phase, see
phase diagram 
in Fig.3.(a), and
pushing up the critical disorder needed for localisation.
The critical disorder now depends on $J'$ and increases 
 from $\Delta \approx  16.5$, at $J'=0$, to   
$\Delta \sim 27$ at $J' \sim 8$, 
before dropping to $\Delta \sim 12$ for
$J' \rightarrow \infty$ (see  Fig.3.(a)).
The optical conductivity data, Fig.3.(b) illustrates
how the low frequency conductivity  {\it increases}
as $J'$ grows from $0.2$ to $2.0$, at $\Delta = 16$, and 
drastically
reduces again for $J' = 20$.
{\it There is an insulator-metal-insulator 
transition with increasing $J'$ for
$26 >  \Delta > 16.5$.}

While the  IMT driven by `small $J'$' 
may be motivated as the `extrapolation'
of the antilocalising effect of spin flip scattering, new 
physical effects emerge rapidly as $J'$ is increased, staying in
the large $\Delta$ regime. 
The reducing resistivity, Fig.1.(a), and Fig.2.(a), exhibits
a minima, and then {\it rises} again with increasing
$J'$. Such behaviour can be viewed as an extension of
the $J'^2$ term seen at weak disorder, but it is probably more
fruitful to approach the problem from the strong coupling,
double-exchange, end.
\begin{figure}

\hspace{1.5cm}
{\epsfxsize=11cm, \epsfysize= 6.0cm, \epsfbox{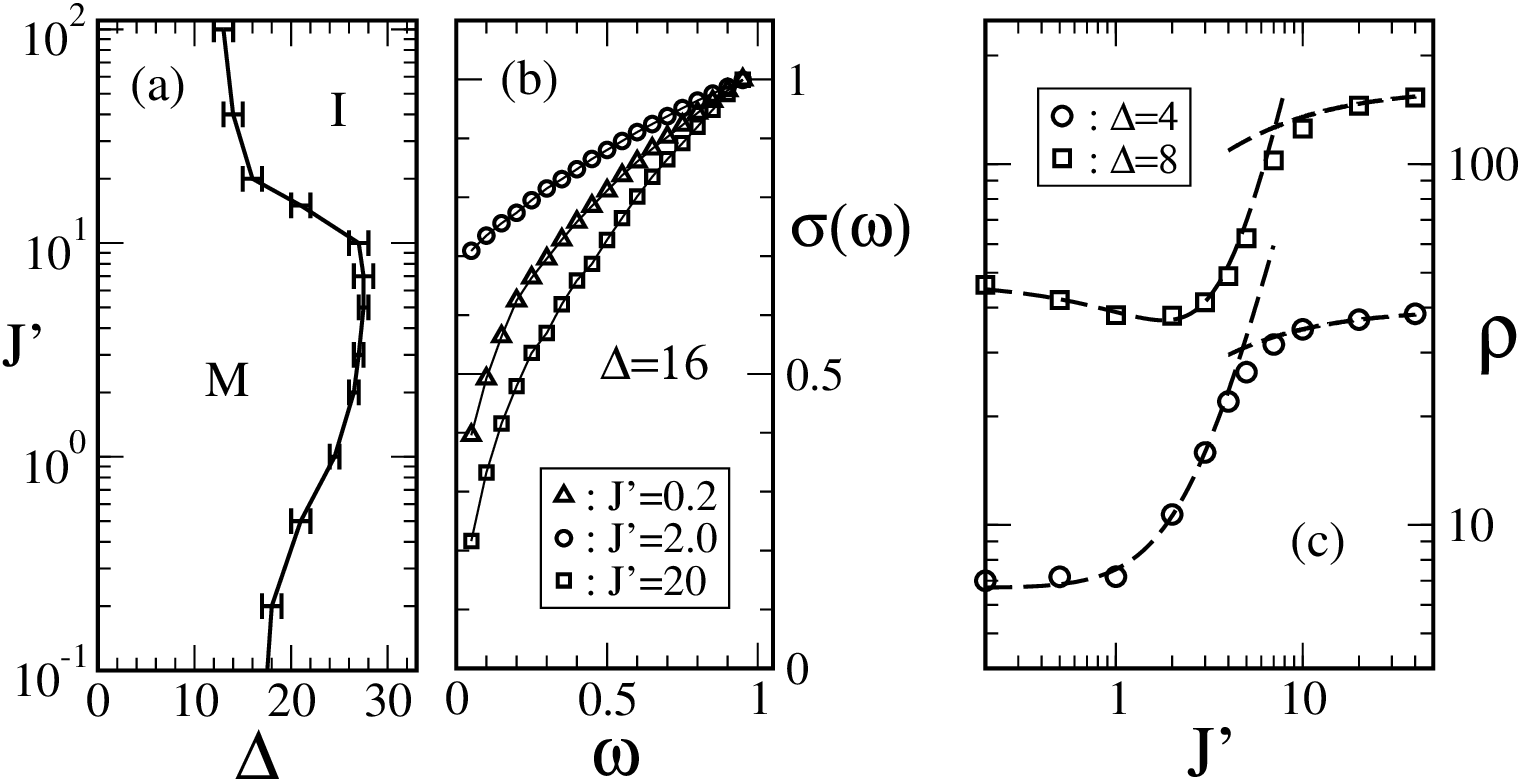}}
\caption{
$(a)$. The insulator-metal phase diagram in $\Delta - J'$,
$(b)$
the optical conductivity illustrating the `I-M-I' feature,
$(c)$ Combining the weak coupling and strong coupling expansions,
Fig.2.(a) and Fig.2.(e), for
a `global' description of the resistivity.
}
\vspace{-2mm}
\end{figure}

$(iv)$~{\it Double exchange with disorder:}
This  regime corresponds to $J'/W \rightarrow \infty$, 
in the presence of  arbitrary structural disorder,
{\it i.e}, the right edge of panel 1.(b).
The form of the 
resistivity $\rho(J')$ arising from `magnetic disorder' 
at large
$J'$ is very different from what one observes in $\rho(\Delta)$
at large $\Delta$. This is because $J'$ contributes to both
`band splitting' and effective disorder, and the effective
disorder saturates as $J'/W \rightarrow \infty$ with 
$J'$ controlling only the band splitting.

A standard transformation \cite{dex-transf}
and projection  maps on the $t-J'$ problem to
a spinless fermion model with hopping amplitudes $t_{ij}$
dependent on nearest neighbour spin orientation:
$ H = \sum_{ij} t_{ij} (\theta, \phi) 
\gamma^{\dagger}_i \gamma^{~}_j +
\sum_i \epsilon_i \gamma^{\dagger}_i \gamma^{~}_i 
 \equiv
\sum_{ij} t_0 
\gamma^{\dagger}_i \gamma_j +
\sum_{ij} \delta t_{ij} \gamma^{\dagger}_i \gamma_j +
\sum_i \epsilon_i \gamma^{\dagger}_i \gamma_i$
with the hopping amplitude specified by
$t_{ij}=-t(cos{\theta_i\over 2} cos{\theta_j \over 2} 
+ sin{\theta_i \over 2} sin{\theta_j \over 2} e^{i(\phi_i - \phi_j )})$. We 
have
split this into the `mean hopping', $t_0 = \langle \vert t_{ij} \vert
\rangle $, and
the fluctuation
$\delta t$. 

The localisation properties
of this model have been studied by Li {\it et al.} \cite{li-dex-dis}, 
although 
they did not calculate the resistivity.
The `hopping disorder' by itself localises less than
$0.5 \%$ of the states in the band, and,
 as we observe, the resistivity
at band center remains finite, $\sim 0.2 \rho_{Mott}$.
On adding structural disorder
the mobility edge moves inward, with localisation of the
full band  occuring at 
$\Delta/t \sim 11.5$.
The critical disorder, visible approximately at large 
$J'$ in Fig.3.(a), can be motivated by the band narrowing
in $t_0$ compared to $t$. In the fully spin disordered phase 
$t_0/t \sim 2/3$, and a crude
scaling of $\Delta_c(J'=0)$ would suggest, $\Delta_c(\infty)/\Delta_c(0)
\sim 2/3$, so that the double exchange model in the paramagnetic
phase would localise at 
$\Delta \sim 11$. The small deviation from this value
 arises from the presence
of $\delta t_{ij}$ and the Berry phase  involved in it.
Apart from the mobility edge calculation \cite{li-dex-dis}
we do not know of any exact results on transport  
in the strongly disordered 
double exchange model.

The growth in resistivity with $\Delta$, remaining at large $J'$,
is visible in the right edge of Fig.1.(b). 
Within our calculation the 
MIT at large $J'$ occurs slightly above $\Delta = 12$. 

$(v)$~{\it Strongly disordered, intermediate coupling metal:}
This is the most complicated regime in the problem and also
the most relevant for real disordered magnetic systems. The
regime corresponds to $\Delta/W \sim 1$ and $J'/W \sim 1$, and,
as obvious from Fig.1, $\rho/\rho_{Mott} > 1$.
This is a nominally `metallic' but diffusive, highly resistive regime.
There are no analytical tools for directly estimating transport properties
in this regime, but much of the physics can be motivated by
an expansion about the double exchange limit.

When $J'$ is finite, both the electron spin states, polarised parallel and
anti-parallel to the core spin, need to be retained. In terms of these states 
the Hamiltonian assumes the form:
$ H = \sum_{ij} t^{\alpha \beta}_{ij} \gamma^{\dagger}_{i \alpha} 
\gamma_{j \beta}
+ \sum_i \epsilon_i n_i 
- {J' \over 2} \sum_i (n_{i \alpha} - n_{i \beta})
$.
At large $J'$ the chemical potential will be in the lower
band, if we want $n < 1$. 
We can split the $t_{ij}^{\alpha \beta}$, as we 
did for the 
single band model, into mean amplitudes and fluctuations. 
The major source of disorder is still $\epsilon_i$, with
additional contribution from the $\delta t_{ij}^{\alpha \beta}$.
The orbital mixing effect of 
`off diagonal' couplings, either in terms of mean amplitude
or fluctuations, is regulated  by the large energy denominator
$J'$.
Although the `reference' problem, $J' \rightarrow \infty$, is
not analytically tractable in the presence of structural disorder,
we expect  orbital mixing
to generate corrections to conductivity $ \sim {\cal O}(1/J')$.
Fig.2.(e)  fits the data, for $\Delta$ upto $8$ and
$J'$ down to $6$, to the form $
\sigma(J', \Delta)~\vert_{J' \rightarrow \infty}
\approx \sigma(\infty, \Delta ) + d_1/J'
$.
Fig.2.(f)  show $\rho(\infty, \Delta)
=1/\sigma(\infty, \Delta)$ and Fig.2.(g) is 
the coefficient $f_1$.

Fig.3.$(a)$~puts together the phase diagram that emerges from 
tracking the conductivity across the full $\Delta-J'$ range.
While the $\Delta$ dependence seems intuitive, in that increasing
structural disorder always leads to a MIT, the effect of $J'$,
at fixed large $\Delta$, is nontrivial. We can identify a 
clear I-M-I transition which, we believe, is a new result.
As discussed before, the $\sigma(\omega)$ in  panel 3(b) 
confirms the I-M-I transition
seen in the resistivity.

Fig.3.(c) puts together the weak coupling and strong coupling fits to
$\rho(J', \Delta)$, for $\Delta < 8$, comparing with the
full numerical data.
The quality of the fit indicates that the dominant transport regimes
in this problem can be classified as we had suggested earlier
and  some of the physics can be motivated by extending concepts 
that are already available in the transport literature.
The regions of MIT cannot be described by these expansions
and should probably 
be accessed  via a complementary scaling approach.

\section{Conclusions}
We have presented exact results on electron transport 
in the background of arbitrary structural and spin disorder
and provided a framework within which the data can be
analysed. We have identified the distinct transport regimes in the model,
explored the Anderson transition in the presence of weak spin disorder, and
located a novel insulator-metal-insulator transition driven by 
increasing magnetic coupling.
In addition to the  intrinsic interest as a non trivial
disorder problem, our results 
should be useful in understanding transport 
in the paramagnetic phase in several currently interesting
magnetic materials.

{\it Acknowledgement} We acknowledge use of the Beowulf cluster at H.R.I.

{}

\end{document}